\documentstyle[aps]{revtex}
\begin{document}
\draft
\title{ Capacitance of  Gated  $\rm \bf  GaAs/Al_{x}Ga_{1-x}As$
Heterostructures Subject to In-plane Magnetic Fields }

\author{ T. Jungwirth,  L. Smr\v{c}ka}
\address{{ \it Institute of Physics, Acad. of Sci. of Czech Rep.,} \\
{\it Cukrovarnick\'{a} 10, 162 00 Praha  6, Czech Republic}}

\date{Received \today}
\maketitle

\begin{abstract}
A detailed analysis of the capacitance of gated $\rm
GaAs/Al_{x}Ga_{1-x}As$ heterostructures is presented. The
nonlinear dependence of the capacitance on the gate voltage and
in-plane magnetic field is discussed together with the
capacitance quantum steps connected with a population of higher
2D gas subbands. The results of full self-consistent numerical
calculations are compared to recent experimental data.
\end{abstract}


In modulation doped $\rm GaAs/Al_{x}Ga_{1-x}As$ heterostructures
a charged channel is formed in GaAs near the interface due to the
electron transfer from donors in $\rm Al_{x}Ga_{1-x}As$. If
a metal Schottky gate is evaporated on the $\rm
Al_{x}Ga_{1-x}As$ surface the electron density $N$ of the GaAs
inversion layer can be controlled by the gate voltage $V_g$. In
strong inversion and at low temperatures the impurity depletion
charges are essentially fixed and the charge induced by the
variation of the gate voltage goes entirely into the inversion
layer. In this regime the heterostructure behaves like
a capacitor with one metal electrode and the second electrode
represented by the channel in GaAs.

The differential capacitance (per unit area) of a gated
structure, $C = d|e|N/dV_g$, is traditionally \cite{i},
\cite{1} written in the form

\begin{equation}
\label{a}
C^{-1} = C_b^{-1} + \frac{\gamma \overline{z}}{\epsilon} +
\frac{1}{e^2 g}.
\end{equation}

\noindent Here $C_b = \epsilon/d_b$ denotes the barrier
capacitance determined by the $\rm Al_{x}Ga_{1-x}As$ barrier
width $d_b$ and the dielectric constant $\epsilon$. The second
term takes into account the distance $\overline{z}$ of the
centroid of the inversion layer charge from the interface. The
numerical prefactor $\gamma$ results from approximating (see
e.g. \cite{ii} and \cite{iii}) the dependence of the subband
energy $E_0$ on $N$ by $dE_0/dN \propto \overline{z}$. In $\rm
GaAs/Al_{x}Ga_{1-x}As$, $\gamma$ lies in the interval 0.5 - 0.7.
The last term of equation (\ref{a}) expresses the dependence of
the capacitance on the density of states $g$ of the 2D electron
gas.

 Recently the capacitance of gated $\rm GaAs/Al_{x}Ga_{1-x}As$
structures has been measured with a high accuracy by Hampton et
al \cite{1}, both as a function of the gate voltage and of the
in-plane magnetic field. The main results of the measurements
are, with a kind permission of authors, reproduced in figure
\ref{f1}. The aim of this paper is to present a detailed analysis
of the data based on the full self-consistent numerical
calculation, going beyond the validity of equation (\ref{a}).

A band diagram of the charged gated structure is sketched in
figure \ref{f2}. Boundary conditions for the electrostatic
potential $\Phi(L_a) = 0$ and $d\Phi/dz(L_a) = 0$ correspond to
the assumption of fixed impurity depletion charges and to
a charge neutrality of the whole structure. The gate voltage
$V_g$ is a difference between the chemical potential $E_F^g/|e|$
of the gate and the chemical potential $E_F/|e|$ of the inversion
layer. At the metallic gate the chemical potential is determined
essentially by the electrostatic potential $\Phi_g$ since the
difference is a constant built-in voltage of the Schottky barrier
$V_B$. We obtain from the Poisson equation

\begin{equation}
\label{b}
\Phi_g = \frac{d_b +  \overline{z}}{\epsilon}|e|N + K \; .
\end{equation}

\noindent The constant $K$ is a contribution to the potential due
to frozen out impurity charges and due to the discontinuity at
the $\rm GaAs/Al_{x}Ga_{1-x}As$ interface. Consequently, the gate
voltage is given by

\begin{equation}
\label{c}
V_g = \frac{E_F}{|e|} + \frac{d_b +
\overline{z}}{\epsilon}|e|N + K + V_B
\end{equation}

\noindent and the reciprocal value of the capacitance  reads

\begin{equation}
\label{d}
C^{-1} = C_b^{-1} + C_c^{-1}
\end{equation}

\noindent where

\begin{equation}
\label{e}
C_c^{-1} = \frac{d}{dN} \left(\frac{\overline{z}}{\epsilon}
N+\frac{E_F}{|e|^2} \right)\; .
\end{equation}

\noindent Thus, the gated structure capacitance can be again
considered as a barrier capacitance $C_b$ and a channel
capacitance $C_c$ in series. The reciprocal value of the channel
capacitance is a sum of two terms: the electrostatic one, related
to the concentration dependent position of the channel centroid
in GaAs and the thermodynamical one originating from the
concentration dependent chemical potential in the semiconductor.
Both these quantities can also depend on the applied magnetic
field.

Up to now a very general analysis of the gated structure
capacitance has been carried out regardless the detail properties
of the inversion layer. In the next step we summarize basic
formulae describing the layer electron structure.

 In a zero magnetic field the standard envelope function
approximation yields the Schr\"{o}dinger equation for
a conduction electron

\begin{equation}
\label{f}
\left(\frac{{\bf
p}^2}{2m}-|e|V_{conf}\left(z\right)-E\right)\psi\left({\bf
r}\right)=0
\end{equation}

\noindent where the confining potential $V_{conf}$ is a sum of
the electrostatic potential $\Phi$, obtained within the Hartree
approximation, and the exchange-correlation term $V_{xc}$,
describing the effects of electron-electron interaction beyond
the Hartree approximation. We use the simple analytic
paramertization of $V_{xc}$ within the local-density-functional
formalism, suggested by Ruden and D\"{o}hler \cite{5}. Thus, the
self-consistent solution to coupled Schr\"{o}dinger equation
(\ref{f}) and Poisson equation for the electrostatic potential
$\Phi$ gives the electron structure of the conduction channel,
including the many-body corrections to $E_0$ and $\overline{z}$
\cite{stern}.

Looking for the eigenfunction of the equation (\ref{f}) in the
form

\begin{equation}
\label{g}
\psi({\bf  r})  =  \frac{1}{\sqrt  S}\exp  (ik_x  x  +  ik_y
y)\phi_n(z)
\end{equation}

\noindent we obtain 2D subbands of allowed energies

\begin{equation}
\label{h}
E(k_x,k_y,n) = \frac{\hbar^2}{2m}(k_x^2+k_y^2) + E_n\; .
\end{equation}

\noindent Equations (\ref{g}) and (\ref{h}) are the mathematical
expression for the separability of the free in-plane
($x,y$-direction) and bound out-of-plane ($z$-direction)
components of the electron motion.

Firstly, let us assume only the lowest subband to be occupied.
Then, due to the parabolic dispersion of the in-plane energy, the
2D density of states per unit area $g$ is a universal constant
$m/\pi\hbar^2$ and the 2D concentration of the electron layer
reads

\begin{equation}
\label{i}
N=g(E_F-E_0)
\end{equation}

\noindent  where $E_0$  is the  lowest energy  of the  bound
out-of-plane motion.

The average position $\overline{z_0}$ of the electron layer is
determined by the $z$-dependent part of the wavefunction as

\begin{equation}
\label{j}
\overline{z_0} = \int z|\phi_0(z)|^2 dz\; .
\end{equation}

Now, the attention can be turned back to the heterostructure
capacitance. Due to (\ref{i}) and (\ref{j}) quasi-analytical
relations for both the thermodynamical and electrostatic parts of
the reciprocal channel capacitance are available. The
electrostatic term is related exclusively to the out-of-plane
component of the electron motion. In the thermodynamical term
both components of the motion play a role. However, according to
(\ref{i}) their influence is independent and we obtain the
channel capacitance in the form

\begin{equation}
\label{l}
C_{c}^{-1}=\frac{\overline{z_0}}{\epsilon} +
\frac{N}{\epsilon}\frac{d\overline{z_0}}{dN} +
\frac{\pi\hbar^2}{m|e|^2} + \frac{1}{|e|^2}\frac{dE_0}{dN} \;.
\end{equation}

Different 2D gas concentrations (or gate voltages) directly yield
different shapes of the confining potential $V_{conf}$ and thus
the bound state eigenenergy $E_0$ is modified. For confining
potentials which do not depart substantially from an exactly
triangular well one obtains the scaling laws $ \overline{z_0}
\propto N^{-1/3}$ and $E_0 \propto N^{2/3}$ \cite {iii}, from
them the relation $dE_0/dN \propto \overline{z_0}$ and the
equation (\ref{a}) is regained. The quantitative deviations can
be expected for more realistic shape of the confining potential
of standard gated structures and, in the case of gated symmetric
wells or inverted structures, the above mentioned scaling law is
not valid at all. Therefore both $dE_0/dN$ and $\overline{z_0}$
have to be determined independently in the course of the
self-consistency procedure.

In magnetic fields parallel to the heterostructure interface the
separability between the in-plane and out-of-plane components of
the channel electron motion is lost. Assuming the magnetic field
${\bf B}\parallel y$ the wavefunction $\phi(z)$ becomes $k_x$
dependent. As discussed in more details in \cite {6,7,8}, this
results in a charge redistribution of the 2D gas and in
a modification of 2D subbands. The centroid $\overline{z_0}$ of
the channel first shifts to the bulk GaAs and then, after the
magnetic field reaches a critical value, it returns back to the
heterostructure interface. The deviation of the energy dispersion
$E(k_x)$ from the parabolic shape yields a density of states
which form depends both on the magnetic field and the electron
concentration \cite{iv}. Finally, the subband edge is shifted to
higher energies by the in-plane field. It follows from the above
mentioned effects on the electron structure that the
electrostatic term and both contributions to the thermodynamical
term are modified by the in-plane magnetic field in a rather
complicated way.

A sufficient amount of information about the heterostructure
capacitance has been given to open the last problem we want to
deal with. Except the very general formula (\ref{e}) previous
conclusions are restricted to 2D electron systems with one
occupied subband. Now we extend the discussion to multi-subband
systems since the filling or depleting of the higher subband is
reflected in the observed capacitance quantum steps.
A heterostructure where the number of occupied subbands is
changed by the gate voltage in the absence of the magnetic field
will be considered. A discussion concerning the in-plane magnetic
field induced depopulation of higher subbands would be analogous.

In the  case of two occupied subbands we obtain the channel
capacitance

\begin{equation}
\label{m}
C_{2c}^{-1}=\frac{\overline{z}}{\epsilon} +
\frac{N}{\epsilon}\frac{d\overline{z}}{dN} +
\frac{\pi\hbar^2}{2m|e|^2} +
 \frac{1}{|e|^2}\frac{d\overline{E}}{dN}
\end{equation}

\noindent in the form similar to (\ref{l}) but with
$\overline{z_0}$ and $E_0$ replaced by

\begin{equation}
\label{n}
\overline{z} = \frac{\overline{z_0}N_0+\overline{z_1}N_1}
{N}\;,
\end{equation}

\begin{equation}
\label{o}
\overline{E} = \frac{E_0+E_1}{2}\;.
\end{equation}

\noindent The quantities $N_0 = (E_F-E_0)\;m/\pi\hbar^2$ and
$N_1 = (E_F-E_1)\;m/\pi\hbar^2$ are the concentrations of
electrons in the first and second subband, $N_0+N_1=N$.

Two terms in equations (\ref{l}), (\ref{m}) do not contain the
derivatives with respect to $N$. The first one, proportional to
the centroid coordinate, does not cause any step when (\ref{l})
is replaced by (\ref{m}) after reaching the critical
concentration $N_c$ since $\overline{z}(N_c+0) =
\overline{z_0}(N_c-0)$. The step in the second term is
a universal constant and reflect the fact that the density of
states is doubled above $N = N_c$. The additional steps come from
the terms with derivatives which can be expressed for $N > N_c$
similarly as for the case of single subband occupancy with a help
of $d\overline{z_i}/dN$, $d\overline{E_i}/dN$, $i = 0, 1$.

The previous analysis shows the principal reasons for the gate
voltage or in-plane magnetic field dependency of the one-subband
channel capacitance as well as for steps in capacitance when
a higher subband is populated. To obtain results comparable to
experimental data full self-consistent numerical calculations,
treating the electron-electron interaction beyond the Hartree
approximation both in zero and non-zero magnetic fields, are
unavoidable since the simple scaling rule leading to (\ref{a}) is
quite unrealistic in the presence of the magnetic field and for
the double subband occupancy.

Figure \ref{f1} presents data of low temperature capacitance
measurements on an ${\rm Al_xGa_{1-x}As/GaAs}$ sample with
$x=0.3$ and channel concentrations from about $1 \times 10^{11}$
to $ 4 \times 10^{11}\,{\rm cm^{-2}}$. Since the shape of the
confining potential depends also on depletion charges parameters
which are not exactly known we performed numerical calculations
for $N_a =4\times 10^{14} cm^{-3}$ and several frozen-in
depletions lengths $L_a$. Due to an unambiguous relation between
the gate voltage dependency of the capacitance, including the
position of the quantum step, and the shape of the quantum well
it is possible to use the experimental curve measured at $B_y =
0$ for determining $L_a$. This way we obtained $L_a=550 \,{\rm
nm}$. (We do not discuss the height of the quantum step since
a current leakage occurred at the gate voltage $V_g=+150 \, {\rm
mV}$, which might quantitatively modify the measured voltage
dependency of the capacitance.) As shown in figure \ref{f3}, if
the theoretical and experimental gate voltage dependencies at the
zero magnetic field are in a good agreement, the theory predicts,
with a very reasonable accuracy, the behaviour of the capacitance
for non zero in-plane magnetic fields. The impurity scattering of
electrons together with the roughness of the interface are
responsible for smearing of the quantum step, we simulate their
influence by introducing the Dingle temperature $T_D \approx 1
{\rm K}$. The effect of the magnetic field on both the
electrostatic and thermodynamical terms in the reciprocal channel
capacitance is shown in figure \ref{f4} to illustrate the
violation of the formula (\ref{a}). Note that the term $1/e^2
\,dE_F/dN$ cannot be simplified due to the field and
concentration dependence of the density of states.

In conclusion, we want to point out that the capacitance
measurements provide a lot of information about 2D electron gas
properties but the data have to be interpreted very carefully. It
is well known that in structures shown in figure \ref{f3} the 2D
channel with higher concentration is confined closer to the $\rm
GaAs/Al_{x}Ga_{1-x}As$ interface resulting in a smaller distance
between electrodes. However, this effect is not reflected
directly in the capacitance due to simultaneous shift of bound
states energies. Similarly, the charge redistribution by the
in-plane magnetic field is accompanied with the shift of subband
edges. Moreover, the modification in the density of states
contributes to the magnetic field induced changes in the
capacitance. Finally, the discontinuities in the 2D density of
states are responsible only partially for the capacitance quantum
steps. We have seen that also the ratio between the filling of
subbands and the relative positions of electron layers in each
subband play an important role.

We are indebted to J.\ P.\ Eisenstein for stimulating discussions
concerning the subject of this paper and for enabling us to use
their experimental data prior to publication. We thank also to
A.\ H.\ MacDonald who turned our attention to this problem. This
work has been supported by the Academy of Science of the Czech
Republic under Grant No. 110414, by the Grant Agency of the Czech
Republic under Grant No. 202/94/1278, by the Ministry of
Education, Czech Republic under contract No. V091 and by NSF,
U.\ S., through the grant NSF INT-9106888.


\begin{figure}
\caption{ Experimental curves of the capacitance as a function of
the gate voltage and of the in-plane magnetic field determined by
Hampton et al \protect \cite{1}.}
\label{f1}
\end{figure}

\begin{figure}
\caption{ Schematic band diagram for a gated
$\rm GaAs/Al_{x}Ga_{1-x}As$ heterostructure.}
\label{f2}
\end{figure}

\begin{figure}
\caption{Self-consistently calculated gate voltage and in-plane
magnetic field dependencies of the relative capacitance. For $B
= 0$ the results for three depletion lengths of acceptors are
presented, $N_a = 4\times 10^{14} cm^{-3}$. Magnetic field
dependencies are plotted only for selected gate voltages and
$L_a=550 \,{\rm nm}$ to make comparison of experimental and
theoretical curves easier.}
\label{f3}
\end{figure}

\begin{figure}
\caption{An increment of the reciprocal channel capacitance
$\delta C_c^{-1}= C_c^{-1}(B) - C_c^{-1}(0)$ (dotted line) in the
magnetic field, decomposed in a thermodynamical term
$1/e^2\,d\delta E_F/dN$ and two contributions
$N/\epsilon\,d\delta\overline{z}/dN$,
$\delta\overline{z}/\epsilon$ to the electrostatic term.}
\label{f4}
\end{figure}

\end{document}